\documentclass[12pt]{iopart}
\usepackage{graphics}
\usepackage{rotating}
\usepackage{epsfig}
\usepackage{color}

\begin{document}
\title{Extracting density-density correlations from \textit{in situ} images of atomic quantum gases}
\author{Chen-Lung Hung, Xibo Zhang, Li-Chung Ha, Shih-Kuang Tung, Nathan Gemelke*, and Cheng Chin}
\address{The James Franck Institute and Department of Physics, The University of Chicago, Chicago, IL 60637, USA\\
$*$Department of Physics, The Pennsylvania State University, 104 Davey Lab, University Park, Pennsylvania 16802}
\ead{clhung@uchicago.edu}

\date{\today}

\begin{abstract}
We present a complete recipe to extract the density-density correlations and the static structure factor of a two-dimensional (2D) atomic quantum gas from \textit{in situ} imaging. Using images of non-interacting thermal gases, we characterize and remove the systematic contributions of imaging aberrations to the measured density-density correlations of atomic samples. We determine the static structure factor and report results on weakly interacting 2D Bose gases, as well as strongly interacting gases in a 2D optical lattice. In the strongly interacting regime, we observe a strong suppression of the static structure factor at long wavelengths. 
\end{abstract}


\maketitle

\section{Introduction}

Fluctuations and correlations result from the transient dynamics of a many-body system deviating away from its equilibrium state. Generally, fluctuations are stronger at higher temperatures and when the system is more susceptible to the external forces (as governed by the fluctuation-dissipation theorem, see \cite{Huang63,Forster75}). Local fluctuations and their correlations can thus be a powerful tool to probe thermodynamics, and to identify phase transition of a many-body system due to the sudden change of the susceptibility to the thermodynamic forces.

Measurement of fluctuations and correlations on degenerate atomic gases can reveal much information about their quantum nature \cite{Altman04}. Experiment examples include the quantum statistics of the atoms \cite{Schellekens05, Folling05,Rom06,Jeltes07,Sanner10}, pairing correlations \cite{Greiner05} and quantum phases in reduced dimensions \cite{Hofferberth08,Manz10}. In these experiments, images of the sample are taken after the time-of-flight expansion in free space, from which the momentum-space correlations are extracted.

\textit{In situ} imaging provides a new and powerful tool to examine the density fluctuations in real space \cite{Esteve06,Gemelke09, Itah10, Muller10, Armijo10,Bakr10,Sherson10, Sanner11,Jacqmin11}, offering a complimentary description of the quantum state. This new tool has been used to resolve spatially separated thermodynamic phases in inhomogeneous samples. From \textit{in situ} measurements, both Mott insulator density plateaus and a reduction of local density fluctuations were observed \cite{Gemelke09, Bakr10, Sherson10, Hung10}. Furthermore, a universal scaling behavior was observed in the density fluctuations of 2D Bose gases \cite{Hung11}.

Precise measurements of spatial correlations, however, present significant technical challenges. In \textit{in situ} imaging, one typically divides the density images into small unit cells or pixels and then evaluates the statistical correlation of the signals in the cells. If both the dimension of the cell and the imaging resolution are much smaller than the correlation length of the sample, the interpretation of the result is straightforward. In practice, because the correlation length of quantum gases is typically on the order of 1~$\mu$m, comparable to the optical wavelength that limits the image resolution, interpreting experimental data is often more difficult. Finite image resolution, due to either diffraction, aberrations or both, contributes to systematic errors and uncertainties in the fluctuation and correlation measurements.

In this paper, we present a general method to determine density-density correlations and static structure factors of quantum gases by carefully investigating and removing systematics due to imaging imperfections. In Section 2, we review the static structure factor and its relation to the real space density fluctuations. In Section 3, we describe how the density fluctuation power spectrum of a non-interacting thermal gas can be used to calibrate systematics in an imperfect imaging system, and show that the measurement can be explained by aberration theory. In Section 4, we present measurements of density fluctuations in weakly interacting 2D Bose gases and strongly interacting gases in a 2D optical lattice, and extract their static structure factors from the density-density correlations.

\section{The density-density correlation function and the static structure factor}

We start by considering a 2D, homogeneous sample at a mean density $\bar{n}$. The density-density correlation depends on the separation $\mathbf{r}_1-\mathbf{r}_2$ between two points, and the static correlation function $\nu(\mathbf{r})$ is defined as \cite{Giogini98}
\begin{eqnarray}
\nu(\mathbf{r}_1-\mathbf{r}_2) &=&  \bar{n}^{-1}\langle \delta n(\mathbf{r}_1) \delta n (\mathbf{r}_2) \rangle \nonumber \\&=& \delta(\mathbf{r}_1-\mathbf{r}_2) + \bar{n}^{-1}\langle \hat{\Psi}^\dag(\mathbf{r}_1) \hat{\Psi}^\dag(\mathbf{r}_2) \hat{\Psi}(\mathbf{r}_1)\hat{\Psi}(\mathbf{r}_2)\rangle - \bar{n}
\end{eqnarray}
\noindent where $\langle ... \rangle$ denotes the ensemble average, and $\delta n(\mathbf{r}) = n(\mathbf{r})-\bar{n}$ is the local density fluctuation around its mean value $\bar{n}$. The Dirac delta function $\delta (\mathbf{r}_1-\mathbf{r}_2)$ represents the autocorrelation of individual atoms, and $\langle \hat{\Psi}^\dag(\mathbf{r}_1) \hat{\Psi}^\dag(\mathbf{r}_2) \hat{\Psi}(\mathbf{r}_1)\hat{\Psi}(\mathbf{r}_2)\rangle = G^{(2)}(\mathbf{r}_1-\mathbf{r}_2)$ is the second-order correlation function \cite{Naraschewski99}. When the sample is completely uncorrelated, only atomic shot noise is present and $\nu(\mathbf{r}_1-\mathbf{r}_2) = \delta (\mathbf{r}_1-\mathbf{r}_2)$. At sufficiently high phase space density, when the inter-particle separation becomes comparable to the thermal de Broglie wavelength $\lambda_{dB}$ or the healing length, density-density correlation becomes non-zero near this characteristic correlation length scale and $\nu(\mathbf{r})$ deviates from the simple shot noise behavior.

The static structure factor is the Fourier transform of the static correlation function \cite{Giogini98, Pitaevskii03}
\begin{eqnarray}
S(\mathbf{k}) &=&  \int \nu(\mathbf{r}) e^{- i \mathbf{k}\cdot\mathbf{r}} d\mathbf{r}, 
\end{eqnarray}
where $\mathbf{k}$ is the spatial frequency wave vector. We can rewrite the static structure factor in terms of the density fluctuation in the reciprocal space as \cite{Pitaevskii03}
\begin{equation}
S(\mathbf{k}) = \frac{\langle \delta n(\mathbf{k}) \delta n(\mathbf{-k})\rangle}{N} = \frac{\langle |\delta n(\mathbf{k})|^2\rangle}{N},
\end{equation}
\noindent where $\delta n(\mathbf{k}) = \int \delta n(\mathbf{r}) e^{-i \mathbf{k}\cdot \mathbf{r}}d\mathbf{r}$, and $N$ is the total particle number. Here, $\delta n(\mathbf{-k}) = \delta n^*(\mathbf{k})$ since the density fluctuation $\delta n(\mathbf{r})$ is real. The static structure factor is therefore equal to the density fluctuation power spectrum, normalized to the total particle number $N$. A non-correlated gas possesses a structureless, flat spectrum $S(\mathbf{k})=1$ while a correlated gas shows a non-trivial $S(\mathbf{k})$ for $k$ near or smaller than inverse of the correlation length $\xi^{-1}$. 

The static structure factor reveals essential information on the collective and the statistical behavior of thermodynamic phases \cite{Forster75, Pitaevskii03,Pines66,Nozieres90}. It has been shown that, through the \emph{generalized fluctuation-dissipation theorem} \cite{Kubo66}, the static structure factor of a Bose condensate is directly related to the elementary excitation energy $\epsilon(\mathbf{k})$ as \cite{Pitaevskii03, Klawunn11}
\begin{equation}
S(\mathbf{k}) = \frac{\hbar^2 k^2}{2 m \epsilon (\mathbf{k})} \coth \frac{\epsilon (\mathbf{k})}{2 k_B T},\label{sfactor}
\end{equation}
where $m$ is the atomic mass, $T$ is the temperature, $\hbar$ is the Planck constant $h$ divided by $2 \pi$, and $k_B$ is the Boltzmann constant. See references~\cite{Pitaevskii03, Pines66, Nozieres90} on the static structure factor of a general system with complex dynamic density response in the frequency domain.

Previous experimental determinations of the static structure factor in the zero-temperature limit, based on two-photon Bragg spectroscopy, have been reported for weakly interacting Bose gases \cite{Kurn99, Steinhauer02} and strongly interacting Fermi gases \cite{Kuhnle10}. Here, we show that $S(\mathbf{k})$ at finite temperatures can be directly determined from \emph{in situ} density fluctuation and correlation measurements.  

\begin{figure}[t]
\begin{center}
\includegraphics[width=0.65\columnwidth,keepaspectratio]{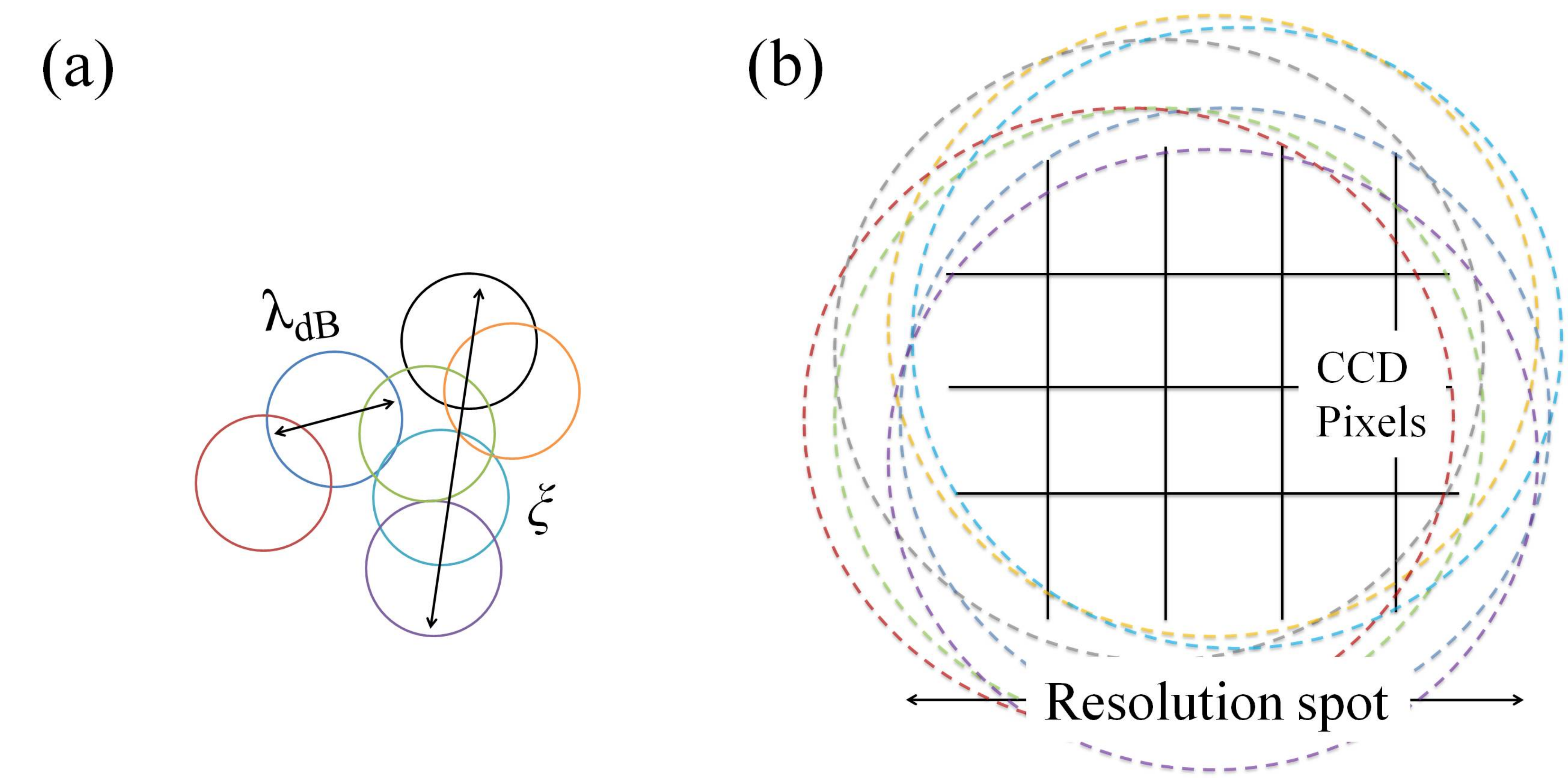}
\caption{A comparison between physical length scales and measurement length scales. In (a), each atom is represented by a color circle with its diameter equal to the thermal de Broglie wavelength, $\lambda_{dB}$. $\xi$ is the correlation length. An ideal measurement detects atom with perfect resolution. (b) shows the experimental condition where the image resolution is larger than the other length scales. Here, the image of an atom forms a resolution limited spot (dashed circles), and is large compared to the correlated area $\xi^2$ and the CCD pixel area. The grid lines represent the CCD pixel array.} \label{fig1}
\end{center}
\end{figure}
 
Experimental determination of $S(\mathbf{k})$ from density fluctuations is complicated by artificial length scales introduced  by the measurement, including, for example, finite image resolution and size of the pixels in the charge-coupled device (CCD). Figure~1 shows a comparison between the measurement length scales (the resolution limited spot size and CCD pixel size), the correlation length $\xi$, and the thermal de Broglie wavelength $\lambda_{dB}$. Ideally, a density measurement should count the atom number inside a detection cell (pixel) with sufficiently high image resolution, and the dimension of the cell should be small compared to the atomic correlation length. In our experiment, the image resolution is determined by the imaging beam wavelength $\lambda=852$~nm, the numerical aperture $\mathrm{N.A.}=0.28$, and the aberrations of the imaging system. The image of a single atom on the CCD chip would form an Airy-disk like pattern with a radius comparable to or larger than $\lambda_{dB}$ or $\xi$.  The imaging magnification was chosen such that the CCD pixel size $\sqrt{A} = 0.66~\mu$m in the object plane is small compared to the diffraction limited spot radius $\sim 1.8~\mu$m. The atom number $N_j$ recorded on the $j$-th CCD pixel is related to the atom number density $\int d \mathbf{r} n(\mathbf{r}) \mathcal{P}(\mathbf{r}_j-\mathbf{r})$, assuming the point spread function $\mathcal{P} (\mathbf{r})$ is approximately flat over the length scale of a single pixel,
\begin{equation}
n_{exp} (\mathbf{r}_j) \equiv \frac{N_j}{A} \approx \int d \mathbf{r} n(\mathbf{r}) \mathcal{P}(\mathbf{r}_j-\mathbf{r}),\label{nexp}
\end{equation}
where $\mathbf{r}_j$ is the center position of the $j$-th pixel in the object plane, $n(\mathbf{r})$ is the atom number density distribution, and the integration runs over the entire $x-y$ coordinate space. The atom number fluctuation measured at the $j$-th pixel is related to the density-density correlation as
\begin{equation}
\langle \delta N_j^2\rangle \approx A^2 \int d \mathbf{r} \int d \mathbf{r}' \langle \delta n(\mathbf{r}) \delta n(\mathbf{r'})\rangle \mathcal{P}(\mathbf{r}_j-\mathbf{r})\mathcal{P}(\mathbf{r}_j-\mathbf{r'}),
\end{equation}
where $\delta N_j = N_j - \langle N_j \rangle$ is the atom number fluctuation around its mean value $\langle N_j \rangle$.

In the Fourier space, Eq.~(5) can be written as
\begin{equation}
\delta n_{exp} (\mathbf{k}_l) \approx  \delta n(\mathbf{k}_l) \mathcal{P}(\mathbf{k}_l),
\end{equation}
where $ \delta n_{exp} (\mathbf{k}_l) \equiv \sum_j \delta N_j e^{-i \mathbf{k}_l\cdot \mathbf{r}_j}$ is the discrete Fourier transform of $\delta N_j$, approximating the continuous Fourier transform. Here, $\mathbf{k}_l=\frac{2\pi}{L}(l_x,l_y)$, $L$ is the linear size of the image, $l_x$ and $l_y$ are integer indices in $\mathbf{k}$-space. From Eq.~(3) and (7), the power spectrum of the density fluctuation is related to the static structure factor as
\begin{equation}
\langle |\delta n_{exp} (\mathbf{k}_l)|^2\rangle \approx N S(\mathbf{k}_l)\mathcal{M}^2(\mathbf{k}_l),
\end{equation}
where the modulation transfer function $\mathcal{M}(\mathbf{k})=|\mathcal{P}(\mathbf{k})|$ accounts for the imaging system's sensitivity at a given spatial frequency $\mathbf{k}$, and is determined by the point spread function. Also, from Eq.~(6), the pixel-wise atom number fluctuation is related to the weighted static structure factor integrated over the $\mathbf{k}$-space,
\begin{equation}
\langle \delta N_j^2\rangle\approx \frac{ \langle N_j \rangle A}{4\pi^2}\int d\mathbf{k} S(\mathbf{k}) \mathcal{M}^2(\mathbf{k}).
\end{equation}

Generalization of the above calculations to arbitrary image resolution and detection cell size is straightforward. In addition to convolving with the point spread function, the measured atom number density must also be convolved with the detection cell geometry. Equation (5) can therefore be written as $N_j/A = \int d\mathbf{k} n(\mathbf{k}) \mathcal{P}(\mathbf{k}) \mathcal{H}(\mathbf{k}) e^{ i \mathbf{k}\cdot\mathbf{r}_j}$, where $\mathcal{H}(\mathbf{k}) = \int_A e^{i \mathbf{k}\cdot\mathbf{r}}d\mathbf{r}/A$ and the integration goes over the area $A$ of the detection cell. This suggests simply replacing $\mathcal{M}^2(\mathbf{k})$ by $\mathcal{M}^2(\mathbf{k})\mathcal{H}^2(\mathbf{k})$ to generalize Eq.~(8) and (9). Finally, in all cases, the discrete Fourier transform defined in Eq.~(7) should well approximate the continuous Fourier transform for spatial frequencies smaller than the sampling frequency $1/\sqrt{A}$. 

We view the factor $\mathcal{M}^2(\mathbf{k})\mathcal{H}^2(\mathbf{k})$ as the general imaging response function, describing how the imaging apparatus responds to density fluctuations occurring at various spatial frequencies. To extract the static structure factor from \textit{in situ} density correlation measurements, one therefore needs to characterize the imaging response function at all spatial frequencies to high precision. Since our pixel-size is much smaller than the diffraction and aberration limited spot size, we will from here forward assume $\mathcal{H}^2(\mathbf{k})=1$; $\mathcal{H}^2(\mathbf{k})$ decays around $k\sim 4/\sqrt{A}=6~\mu$m$^{-1}$ ($1/e$ radius), which is much larger than $k=2\pi \mathrm{N.A.}/\lambda=2.1~\mu$m$^{-1}$, where $\mathcal{M}^2(\mathbf{k})$ terminates.

\section{Measuring the imaging response function $\mathcal{M}^2(\mathbf{k})$}
In this section, we show how to use density fluctuations of thermal atomic gases to determine the imaging response function $\mathcal{M}^2(\mathbf{k})$. 
Other approaches based on imaging individual atoms can be found, for example, in Refs.~\cite{Bucker09,Karski09}.

\subsection{Experiment}
Measuring density fluctuations in low density thermal gases provides an easy way to precisely determine the imaging response function. An ideal thermal gas at low phase-space density has an almost constant static structure factor up to $k=\lambda_{dB}^{-1}$ \cite{Naraschewski99} which, in our case, is larger than the sampling frequency $1/\sqrt{A}$. Therefore, the density fluctuation power spectrum derived from an ideal thermal gas reveals the square of the modulation transfer function, as indicated by Eq.~(8). 

\begin{figure}[b]
\begin{center}
\includegraphics[width=0.85\columnwidth,keepaspectratio]{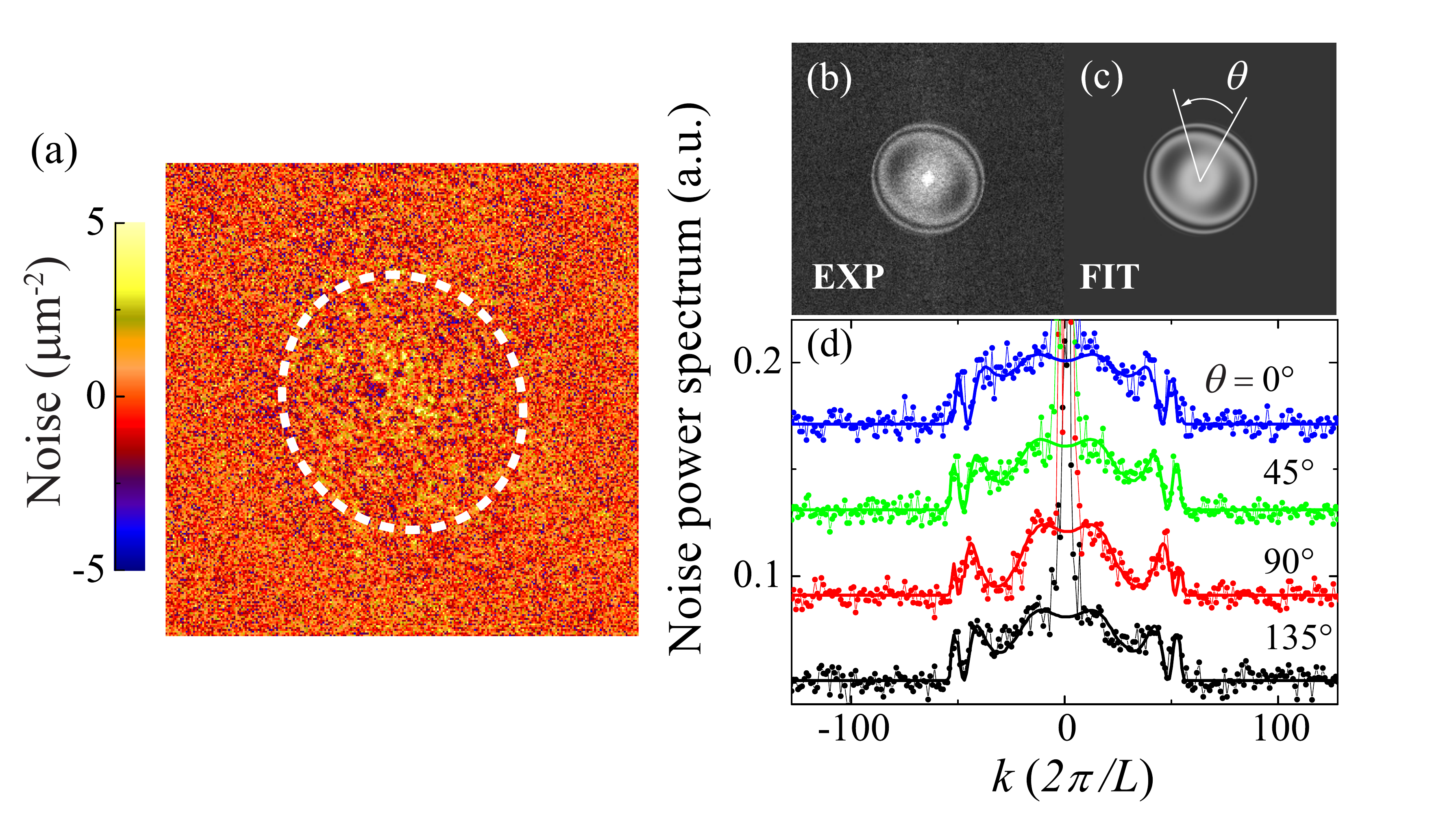}
\caption{Determination of the imaging response function $\mathcal{M}^2(\mathbf{k})$ from \textit{in situ} images of 2D thermal gases. (a) Image noise of a 2D thermal gas. The dashed ellipse encircles the location of thermal atoms. (b) Noise power spectrum evaluated from $60$ images, using discrete Fourier transform defined following Eq.~(7). Zero spatial frequency is shifted to the image center. (c) Fit to the image noise power spectrum using imaging response function defined in Eq.~(\ref{mfit}) and aberration parameters defined in Eqs.~(\ref{pupil}) and (\ref{waveaber}). (d) Sample line-cuts of experiment (circles) and fit (solid lines), with cutting angle $\theta$ indicated in the graph. The profiles are plotted with offset for clarity. Image size: $L^2= 256 \times 256$ pixels.} \label{fig2}
\end{center}
\end{figure}

We prepare a 2D thermal gas by first loading a three-dimensional $^{133}$Cs Bose-Einstein condensate with $2 \times 10^4$ atoms into a 2D pancake-like optical potential with trap frequencies $\omega_z = 2\pi \times 2000~$Hz (vertical) and $\omega_r = 2\pi \times 10~$Hz (horizontal) \cite{Hung10, Hung11}. We then heat the sample by applying magnetic field pulses near a Feshbach resonance. After sufficient thermalization time, we ramp the magnetic field to 17~G where the scattering length is nearly zero. The resulting thermal gas is non-interacting at a temperature $T=90~$nK and its density distribution is then recorded through \emph{in situ} absorption imaging \cite{Hung11}.

We evaluate the density fluctuation power spectrum  $\mathcal{M}^2_{exp}(\mathbf{k}_l) = \langle |\delta n_{exp}(\mathbf{k}_l)|^2\rangle$ using $60$ thermal gas images (size: $256 \times 256$ pixels). Figure~2(a) shows a sample of the noise recorded in the images. Outside the cloud (whose boundary roughly follows the dashed line), the noise is dominated by the optical shot noise, and is therefore uncorrelated and independent of spatial frequency. In the presence of thermal atoms, we observe excess noise due to fluctuations in the thermal atom density. The noise power spectrum is shown as an image in Fig.~2(b), with line-cuts shown in Fig.~2(d). We note that the power spectrum acquires a flat offset extending to the highest spatial frequency, due to the photon shot noise in the imaging beam. Above the offset, the contribution from atomic density fluctuations is non-uniform and has a hard edge corresponding to the finite range of the imaging response function. Close to the edge, ripples in the noise power spectrum appears because of aberrations of the imaging optics, discussed in later paragraphs. Finally, the bright peak at the center corresponds to the large scale density variation due to the finite extent of the trapped atoms, and is masked out in our following analysis.

To fully understand the imaging response function with imaging imperfections, we compare our result with calculations based on Fraunhofer diffraction and aberration theory \cite{Goodma05} as described in the following paragraphs.

\subsection{Point spread function in absorption imaging}
We consider a single atom illuminated by an imaging beam, the latter is assumed to be a plane wave with a constant phase across both the object and the image planes. The atom, driven by the imaging electric field, scatters a spherical wave (dark field) which interferes destructively with the incident plane wave \cite{Ketterle99}. The dark field is clipped by the limiting aperture of the imaging optics and is distorted by the imaging aberrations. An exit pupil function $p$ is used to describe the aberrated dark field at the exit of the imaging optics \cite{Goodma05}, and its Fourier transform $p(\mathbf{k})$ with ${\mathbf{k} \propto \mathbf{R}}$ describes the dark field distribution on the CCD chip, where $\mathbf{R}$ is the position in the image plane. The image of a single atom is then an absorptive feature formed by the interference between the dark field and the incident plane wave in the image plane.

We extend this to absorption imaging of many atoms with density $n(\mathbf{r})= \sum_i \delta(\mathbf{r}-\mathbf{r}_i)$, where $\mathbf{r}_i$ is the location of the $i$-th atom in the object plane. The total scattered field in the image plane is\footnote{We consider the density $n$ of the 2D gas in a range that the photon scattering cross section remains density-independent \cite{Rath10}.} $\Delta E =\sum_i \epsilon p(\mathbf{k}-\mathbf{k}_i)$, with each atom contributing a dark field amplitude $\epsilon$, and $\mathbf{k}$ relates to the position $\mathbf{r}$ in the object plane through $\mathbf{k}=\mathbf{r}/ad$, where $a$ is the radius of the limiting aperture and $d = \lambda / (2 \pi \mathrm{N.A.})$. The dark field $\epsilon \propto e^{i\delta_s} E_0$ is proportional to the incident field $E_0$, and carries with a phase shift $\delta_s$ associated with the laser beam detuning from atomic resonance. For a thin sample illuminated by an incident beam with intensity $I_0$, the beam transmission is $t^2=|E_0+\Delta E|^2/|E_0|^2 \approx 1 + 2 \Re[\Delta E/E_0]$ and the atomic density $n_{exp} \propto -\ln(t^2) + (1-t^2)I_0/I_{sat}$ \cite{Reinaudi07,Hung11} leads to $n_{exp} \propto -2(1+I_0/I_{sat})\Re[\Delta E/E_0] \propto \sum_i \Re [ e^{i\delta_s}p(\mathbf{k}-\mathbf{k}_i) ]$. Here, $\Re[.]$ refers to the real part and $I_{sat}$ is the saturation intensity for the imaging transition. Comparing the above expression to Eq.~(\ref{nexp}), we derive the point spread function as $\mathcal{P}(\mathbf{r}) \propto \Re [e^{i\delta_s}p(\mathbf{k})]|_{\mathbf{k}=\mathbf{r}/ad}$, in contrast to the form $|p(\mathbf{k})|^2$ in the case of fluorescence or incoherent imaging. 

When the dark field passes through aberrated optics, neither the amplitude nor the phase at the exit pupil is uniform, but is distorted by imperfections of the imaging system. To account for attenuation and phase distortion, We can modify the exit pupil function as
\begin{eqnarray}
p(r_p,\theta_p)=U(r_p/a,\theta_p)e^{i \Theta(r_p/a,\theta_p)},\label{pupil}
\end{eqnarray}
where $r_p$ and $\theta_p$ are polar coordinates on the exit pupil, $U(\rho,\theta)$ is the transmittance function, and $\Theta(\rho, \theta)$ is the wavefront aberration function. We assume $U$ to be azimuthally symmetric and model it as $U(\rho)= H(1-\rho)e^{-\rho^2/\tau^2}$, where $H(x)$ is the Heaviside step function setting a sharp cutoff when $r_p$ reaches the radius $a$ of the limiting aperture. The factor $e^{-\rho^2/\tau^2}$, with $1/e$ radius $r_p=a\tau$, is used to model the weaker transmittance at large incident angle due to, e.g., finite acceptance angle of optical coatings. For the commercial objective used in the experiment, we need only to include a few terms in the wavefront aberration function
\begin{equation}
\Theta(\rho,\theta) \approx S_0 \rho^4 + \alpha \rho^2 \cos(2\theta - 2 \phi) + \beta \rho^2 ,\label{waveaber}
\end{equation}
where the parameters used to quantify the aberrations are: $S_0$ for spherical aberration, $\alpha$ for astigmatism (with $\phi$ the azimuthal angle of the misaligned optical axis), and $\beta$ for defocusing due to atoms not in or leaving the focal plane during the imaging.

Using the exit pupil function in Eq.~(\ref{pupil}), we can evaluate the point spread function via $\mathcal{P}(\mathbf{r}) \propto \Re [e^{i\delta_s}p(\mathbf{k})]|_{\mathbf{k}=\mathbf{r}/ad}$ with proper normalization. We can also calculate the modulation transfer function $\mathcal{M}(\mathbf{k})=|\mathcal{P}(\mathbf{k})|$ . In fact, determination of any one of $p(\mathbf{r}_p)$, $\mathcal{P}(\mathbf{r})$, or $\mathcal{M}(\mathbf{k})$ leads to a complete characterization of the imaging system including its imperfections.

\subsection{Modeling the imaging response function}

We fit the exit pupil function $p$ in the form of Eq.~(\ref{pupil}), using a discrete Fourier transform, by comparing 
\begin{equation}
\mathcal{M}^2_{fit}=|\mathcal{FT}(\Re[e^{i\delta_s}\mathcal{FT}(p)])|^2\label{mfit}
\end{equation}
to the thermal gas noise power spectrum $\mathcal{M}^2_{exp}$ shown in Fig.~2(b). Here, $\mathcal{FT}(.)$ denotes the discrete Fourier transform. Figure~2(c) shows the best fit to the measurement, which captures most of the relevant features in the experiment data. Sample line-cuts with uniform angular spacing are shown in Fig.~2(d). This experimental method can in principle be applied to general coherent imaging systems, provided the signal-to-noise ratio of the power spectrum image is sufficiently good to resolve all features contributed by the aberrations. Moreover, one can obtain analytic expressions for the point spread function and the modulation transfer function once the exit pupil function is known (see Appendix A). 

Having determined the imaging response function, one can remove systematic contributions from imaging imperfections to the static structure factor as extracted from the power spectrum of atomic density fluctuations, see Eq.(8). 

\section{Measuring density-density correlations and static structure factors of interacting 2D Bose gases}

We measure the density-density correlations of interacting 2D Bose gases based on the method presented in the previous sections. This study is partially motivated by a finding in our earlier work that the local density fluctuation of a 2D Bose gas is suppressed when it enters the Berezinskii-Kosterlitz-Thouless (BKT) fluctuation and the superfluid regions \cite{Hung11}. We attributed this phenomenon to the emergence of long density-density correlation length exceeding the size of the imaging cell and the resolution. This results in a smaller pixel-wise fluctuation $\delta N^2/A$ than the simple product of the thermal energy $k_B T$ and the compressibility $\kappa$, as is expected from the classical fluctuation-dissipation theorem (FDT) \cite{Huang63}. Below, we present a careful analysis of the density-density correlations of interacting 2D Bose gases and discuss the role of correlations in the FDT. 

\begin{figure}[t]
\begin{center}
\includegraphics[width=0.6\columnwidth,keepaspectratio]{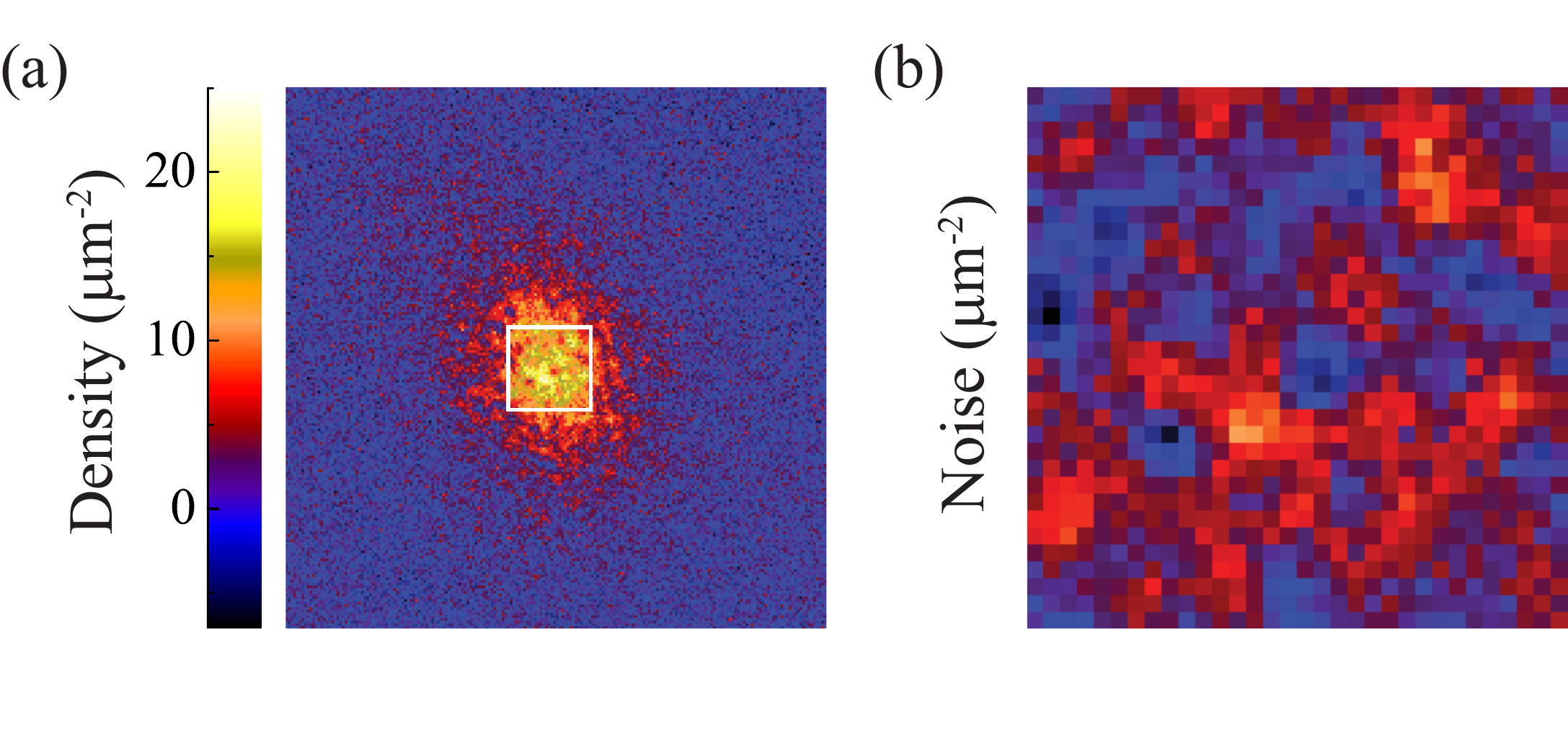}
\caption{Illustration of the patch selected for the static structure factor analysis. (a) shows a typical cloud image of $200\times 200$ pixels. The selected patch is located at the center of the cloud, bounded by a box with an area of $32\times 32$ pixels. (b) shows the density fluctuations inside the patch.} \label{fig_patch}
\end{center}
\end{figure}

To extract local properties from a trapped sample, we limit our analysis to a small central area of the sample where the density is nearly flat. In addition, the area is chosen to be large enough to offer sufficient resolution in the Fourier space. We choose the patch size to be 32$\times$32 pixels. Figure~\ref{fig_patch} shows a typical image and the density fluctuations inside the patch.

\begin{figure}[t]
\begin{center}
\includegraphics[width=0.7\columnwidth,keepaspectratio]{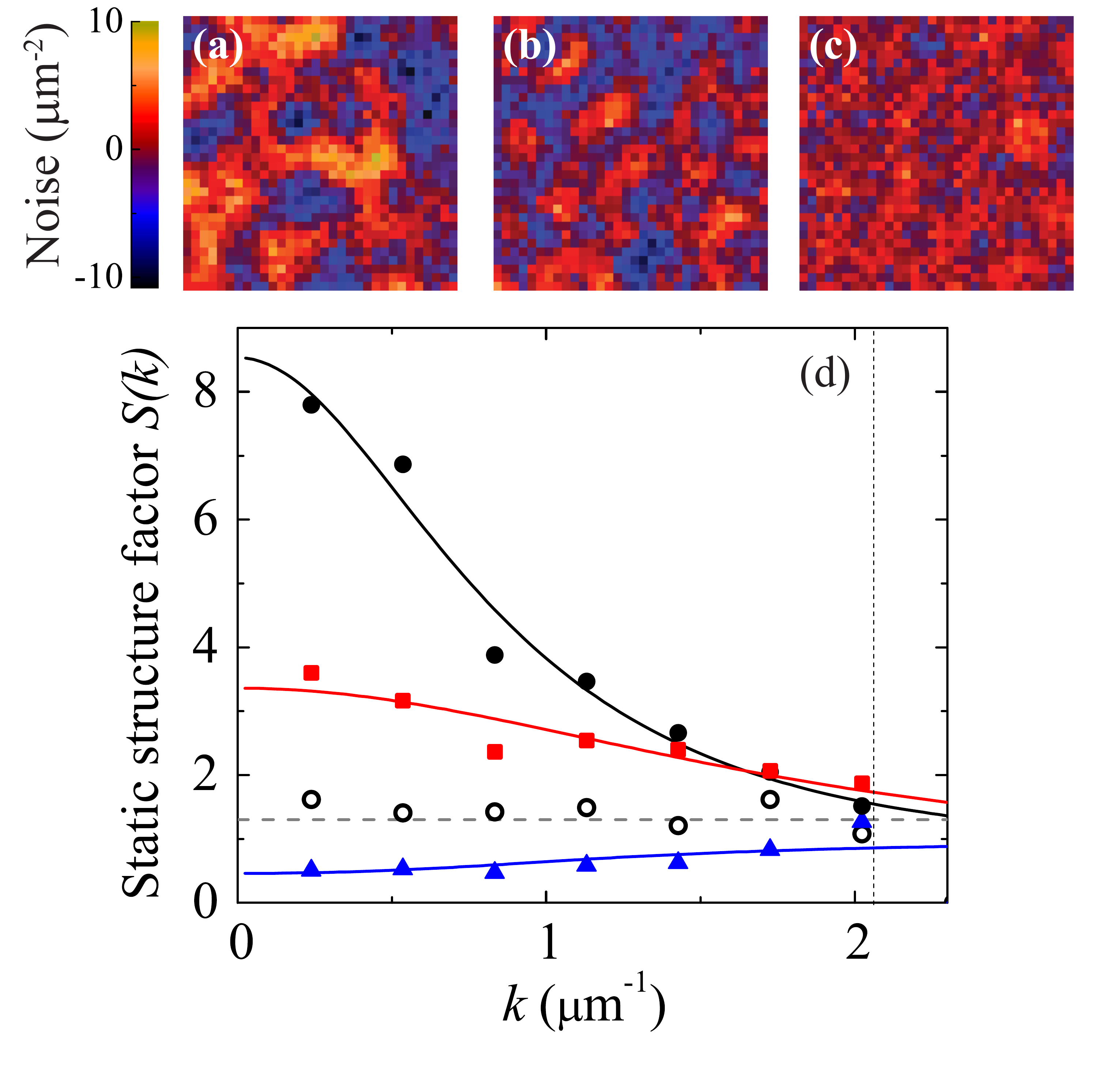}
\caption{Density fluctuations and the static structure factors of 2D Bose gases. (a) and (b) Image noise of weakly interacting 2D Bose gases in the superfluid phase at dimensionless interaction strength $g=0.05$ and 0.26. (c) Image noise of a strongly interacting 2D Bose gas at $g_\mathrm{eff}=1.0$ prepared in a 2D optical lattice at a depth of 7~$E_R$. (d) shows the static structure factors extracted from the noise power spectra of interacting 2D gases as shown in (a) (black circles), (b) (red squares), and (c) (blue triangles). The static structure factor of an ideal thermal gas at phase space density $n\lambda^2_{dB}=0.5$ (open circles) and the expected value of $S(k) \approx 1.3$ (gray dashed line) are plotted for comparison. Solid lines are the guides to the eye. Vertical dashed line indicates the resolution limited spatial frequency $k=2\pi \mathrm{N.A.}/\lambda = 2.1~\mu$m$^{-1}$.} \label{fig_sfactor}
\end{center}
\end{figure}

To ensure that we obtain an accurate static structure factor using the small patch, we perform a measurement on a non-interacting 2D thermal gas at a phase space density $n\lambda_{dB}^2=0.5$ and compare the measured static structure factor to the theory prediction \cite{Naraschewski99}. We first calculate the imaging response function $\mathcal{M}^2(\mathbf{k})$ for a patch size of $32 \times 32$~pixels and divide the thermal gas noise power spectrum by $\mathcal{M}^2(\mathbf{k})$. The resulting spectrum should represent the static structure factor of an ideal 2D thermal gas. In Fig.~\ref{fig_sfactor}, we plot the azimuthally averaged static structure factor with data points uniformly spaced in $k$, up to the resolution limited spatial frequency $k = 2\pi \mathrm{N.A.}/\lambda$. The measured static structure factor is flat and agrees with the expected value of $S(k) \approx 1.3$ for $k<\lambda_{dB}^{-1}=2~\mu$m$^{-1}$\footnote{Following the calculation in Ref.~\cite{Naraschewski99}, we find the static correlation function of an ideal 2D thermal gas is $\nu(r)=\delta(r)+|g_1(z,e^{-\pi r^2/\lambda_{dB}^2})|^2/g_1(z,1)\lambda^2_{dB}$, where $z=e^{\mu/k_B T}$ is the local fugacity and $g_\gamma(x,y)=\sum_{k=1}^\infty x^ky^{1/k}/k^\gamma$ is the generalized Bose function. Fourier transforming $\nu(r)$ to obtain the static structure factor $S(k)$, we find $S(k)\approx1.3$ remains flat for $k \lambda_{dB}<1$.}. 

Applying the same analysis to \emph{interacting} 2D Bose gases, we observe very different strengths and length scales for the density fluctuations. In Fig.~\ref{fig_sfactor}(a-c), we present the single-shot image noise of samples prepared under three different conditions: weakly interacting gases at the temperature $T=40~$nK (below the BKT critical point), with dimensionless interaction strength\footnote[1]{The dimensionless interaction strength of a weakly interacting 2D Bose gas is $g=\sqrt{8 \pi} a_s/l_z$, where $a_s$ is the atomic scattering length and $l_z=\sqrt{\hbar/m\omega_z}$ is the vertical harmonic oscillator length. For a 2D gas in a 2D optical lattice, the effective  interaction strength is $g_\mathrm{eff}= mU l^2/\hbar^2$ \cite{Zhang11}, where $U$ is the on-site interaction and $l$ is the lattice constant.} $g = 0.05$ and 0.26 (phase space density $n\lambda^2_{dB}=9$ and $7$); and a strongly interacting 2D gas at the temperature $T=8~$nK, prepared in a 2D optical lattice at a mean site occupancy number of $2.6$, and a depth of $7~E_R$, where $E_R=h \times 1.3$~kHz is the recoil energy. Due to the tight confinement, the sample in the optical lattice has a high effective interaction strength\footnotemark[1] $g_\mathrm{eff} = 1.0$ \cite{Zhang11}. Details on the preparation of the 2D gases in the bulk and in an optical lattice can be found in Refs.~\cite{Hung11} and \cite{Zhang11}, respectively. 

The difference in the density fluctuations shown in Fig.~\ref{fig_sfactor}(a-c) can be characterized in their static structure factors shown in Fig.~\ref{fig_sfactor}(d). 
We observe positive correlations above the shot noise level $S(k)=1$ in the two weakly interacting samples. The one at $g = 0.05$ shows stronger density correlations at small $k$ than does the sample at $g = 0.26$. The enhanced density correlations $S(k) > 1$ at low momenta are expected since the thermally induced phonon excitations can populate states with length scale $1/k$ longer than the healing length $\xi = 1/\sqrt{ng}$. For gases with stronger interactions, excitations cost more energy and the excited states are less populated. At smaller $g$, the correlation length is longer and, therefore, the static structure factor decays at a smaller $k$.

The most intriguing observation is the negative correlations $S(k) <1$ in the strongly interacting gas with $g_\mathrm{eff}=1.0$. We observe a below-shot-noise spectrum at low momentum $k$, showing that long wave-length excitations are strongly suppressed due to a stronger interaction energy $n\hbar^2g_\mathrm{eff}/m = k_B \times 34 $~nK compared to the thermal energy $k_B \times 8 $~nK. As the momentum $k$ increases, the excitation populations gradually return to the shot noise level. Our observation is consistent with the prediction in Ref.~\cite{Klawunn11} that when the thermal energy drops below the interaction energy, global density fluctuations in a superfluid are suppressed. 

Finally, we discuss the contribution of finite density-density correlations in the FDT. Including correlations, we can write the FDT as $k_B T \kappa(\mathbf{r}) =  \int \langle \delta n(\mathbf{r}) \delta n(\mathbf{r}')\rangle d\mathbf{r}' = n(\mathbf{r})S(0)$ \cite{Klawunn11}.  We compare the measured static structure factor, extrapolated to zero-$k$, to the value of $k_B T \kappa / n$, where $\kappa = \partial n /\partial \mu$ is the experimentally determined compressibility \cite{Hung11}, and indeed find that $n S(0)$ equals to $k_B T \kappa $ to within our experimental uncertainties of $10\sim 20\%$ for all three interacting samples. This agreement shows that the measured correlations and thus the static structure factor can be linked to the thermodynamic quantities via the FDT. Our ability to determine $S(0)$ and $\kappa$ from \emph{in situ} images also suggests a new scheme to determine temperature of the sample from local observables as $k_B T=nS(0)/\kappa$.

\section{Conclusion}
We demonstrated the extraction of density-density correlations and static structure factors from \emph{in situ} images of 2D Bose gases. Careful analysis and modeling of the imaging response function allow us to fully eliminate the systematic effect of imaging imperfections on our measurements of density-density correlations. For thermal gases, our measurement of the static structure factor agrees well with theory. For interacting 2D gases below the BKT critical temperature, intriguingly, we observe positive density-density correlations in weakly interacting samples ($g \ll 1$) and negative correlations in the strong interaction regime ($g_\mathrm{eff} = 1.0$). For all interacting gases, our static structure factor measurements agree with the prediction of the FDT as $S(0)=k_BT \kappa/n$. Extension of our 2D measurement can further test the prediction of anomalous density fluctuations in a condensate \cite{Giogini98,Astrakharchik07,Meier99, Zwerger04} and strong correlations in the quantum critical region \cite{Sachdev99,Sachdev06}. Finally, our analysis can be applied to perform precise local thermometry \cite{Zhou09} and can potentially be used to extract the local excitation energy spectrum through the application of the generalized fluctuation-dissipation theorem \cite{Pitaevskii03,Kubo66}.

\section{Acknowledgement}
We thank C.-C. Chien, D.-W. Wang, Q. Zhou, and T.-L. Ho for useful discussions. This work was supported by NSF (grant numbers PHY-0747907, NSF-MRSEC DMR-0213745), the Packard foundation, and a grant from the Army Research Office with funding from the DARPA OLE program.

\appendix
\section{Full analysis of the point spread function and the modulation transfer function} 
\subsection{Point spread function}
Here, we describe our approach to characterize imaging imperfections using extended Nijboer-Zernike diffraction theory \cite{Janssen02}. To obtain the point spread function from the exit pupil function $p(r_p,\theta_p)$, it is convenient to first decompose the exit pupil function using a complete set of orthogonal functions on the unit disk in the polar coordinates
\begin{eqnarray}
p(r_p,\theta_p)= \sum_{n=0}^{\infty}\sum_{m=- n}^n \beta_n^m Z^m_n(\frac{r_p}{a},\theta_p),\label{pupilexpand}
\end{eqnarray}
\noindent where $Z^m_n (\rho, \theta) = R^{|m|}_n(\rho) e ^{i m \theta}$ is a Zernike polynomial, the radial function $R^m_n(\rho)=\sum^{(n-m)/2}_{k=0}\frac{(-1)^k(n-k)!}{k![(n+m)/2-k]![(n-m)/2-k]!}\rho^{n-2k}$ terminates at $\rho=1$, and $R^m_n = 0$ when $n-m$ is odd. The expansion coefficient $\beta_n^m$ is given by
\begin{equation}
\beta_n^m = \frac{n+1}{\pi a^2} \int_0^a\int_0^{2\pi} p(r_p,\theta_p) Z_n^{-m} (r_p/a,\theta_p) r_p d r_p d\theta_p.\label{zbcoeff}
\end{equation}
If we then apply the expansion Eq.~(\ref{pupilexpand}) to the Fourier transform of the exit pupil function $p(k,\theta)=\int_0^a\int_0^{2\pi}p(r_p,\theta_p)e^{-i k r_p \cos(\theta-\theta_p)}r_p dr_p d\theta_p$ and carry out the integration using the Zernike-Bessel relation $\int_0^1 R_n^m(\rho)J_m(\xi\rho)\rho d\rho=(-1)^{(n-m)/2}J_{n+1}(\xi )/\xi $,
we arrive at the following formula
\begin{eqnarray}
p(k,\theta)= 2\pi a^2 \sum_{n=0}^{\infty}\sum_{m=- n}^n i^n \beta_n^m  e^{im(\theta+\pi)}\frac{J_{n+1}(ka)}{ka},\label{ftp}
\end{eqnarray}
where $J_n(z)$ is the $n$-th order Bessel function of the first kind. The point spread function $\mathcal{P}(r,\theta)$ is the real part of $e^{i\delta_s}p(k,\theta)|_{k=r/ad}$ with proper normalization
\begin{eqnarray}
\mathcal{P}(r, \theta)&=& \frac{1}{\mathcal{N}}\sum_{n=0}^{\infty}\sum_{m=- n}^n \Re[i^n\beta_n^m e^{im(\theta+\pi)+i\delta_s}]\frac{J_{n+1}(r/d)}{r/d},\label{psf}
\end{eqnarray}
where $\mathcal{N}=2\pi d^2 \Re[e^{i\delta_s}p(\mathbf{r}_p)]|_{\mathbf{r}_p=0}=2 \pi d^2 \cos \delta_s$ is the normalizing factor such that $\int \mathcal{P}(\mathbf{r}) d^2r =1$. For a unaberrated system, only $\beta_0^0$ is non-zero and the above equation reduces to the form $J_1(z)/z$, as expected. 

\begin{figure}[t]
\begin{center}
\includegraphics[width=0.95\columnwidth,keepaspectratio]{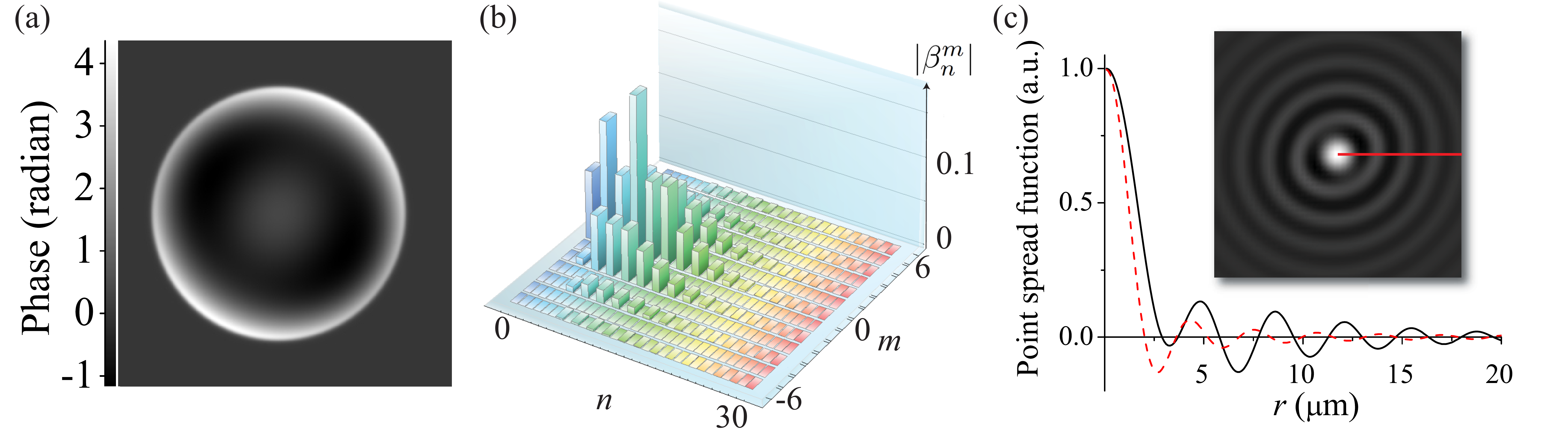}
\caption{Analysis of the imaging aberrations and the point spread function. (a) shows the wavefront aberration (the phase of the exit pupil function) determined from the fit to the experiment. (b) shows the expansion coefficients $|\beta_n^m|$ determined from Eq.~(\ref{zbcoeff}), using the exit pupil function in (a). (c) shows a line-cut of the derived point spread function (solid line). The unaberrated point spread function is plotted for comparison (dashed line). Inset shows the 2D distribution of the aberrated point spread function and red line indicates the direction of the line-cut. Image size is ($33~\mu$m)$^2$. } \label{fig3}
\end{center}
\end{figure}

Using the above equations, we can derive the point spread function from the fitted exit pupil function shown in Fig.~\ref{fig3}(a). We calculate the expansion coefficients $\beta_n^m$ and evaluate the corresponding point spread function, see Fig.~\ref{fig3}(b-c). 

\subsection{Modulation transfer function}

It is straightforward to evaluate the modulation transfer function $\mathcal{M}(\mathbf{k})=|\mathcal{P}(\mathbf{k})|$ and the imaging response function $\mathcal{M}^2(\mathbf{k})=|\mathcal{P}(\mathbf{k})|^2$ directly from the exit pupil function $p(r_p,\theta_p)$. Since the point spread function can be written as $\mathcal{P}(\mathbf{r})=[e^{i\delta_s}p(\mathbf{k})+e^{-i\delta_s}p^*(\mathbf{k})]/4\pi a^2\mathcal{N}|_{\mathbf{k}=\mathbf{r}/ad}$, its Fourier transform is
\begin{eqnarray}
\mathcal{P}(\mathbf{k})&=&\frac{\pi d^2}{\mathcal{N}}  [e^{i\delta_s}p(r_p,\theta+\pi)+e^{-i\delta_s}p^*(r_p,\theta)]|_{r_p=kad}, \label{otf}
\end{eqnarray}
where $k=|\mathbf{k}|$ is the spatial frequency and $\theta$ is the polar angle of $\mathbf{k}$ in the image plane. From Eq.~(\ref{otf}), the imaging response function is $\mathcal{M}^2(\mathbf{k}) \propto |p(kad,\theta+\pi)+e^{-2i\delta_s}p^*(kad,\theta)|^2$. This result shows that the phase shift $\delta_s$ is important since $\mathcal{M}^2(\mathbf{k})$ depends on the interference between $p(kad,\theta+\pi)$ and $p^*(kad,\theta)$. The transmittance $U$, defined in the exit pupil function Eq.~(\ref{pupil}), accounts for the radial envelope in $\mathcal{M}^2(\mathbf{k})$, leading to the sharp edge at $k = d^{-1}=2\pi \mathrm{N.A.}/\lambda$. Either the continuous function Eq.~(\ref{otf}) or the discrete Fourier model Eq.~(\ref{mfit}) can be used to calculate the imaging response function.

\end{document}